\documentclass[fleqn,10pt]{wlscirep}
\usepackage[utf8]{inputenc}
\usepackage[T1]{fontenc}
\title{HCP Multi-Pipeline: a derived dataset to investigate analytical variability in fMRI}

\author[1,*]{Elodie Germani}
\author[2]{Elisa Fromont}
\author[1,$\dag$]{Pierre Maurel}
\author[1,$\dag$,*]{Camille Maumet}
\affil[1]{Univ Rennes, Inria, CNRS, Inserm, IRISA UMR 6074, Empenn ERL U 1228, F-35000 Rennes, France}
\affil[2]{Univ Rennes, IUF, Inria, CNRS, IRISA UMR 6074, F-35000 Rennes, France}

\affil[*]{corresponding author(s): Camille Maumet (camille.maumet@inria.fr), Elodie Germani (elodie.germani@irisa.fr)}

\affil[$\dag$]{Joint senior authorship}

\begin{abstract}
Results of functional Magnetic Resonance Imaging (fMRI) studies can be impacted by many sources of variability including: different sampling strategies for the participants, different acquisition protocols and material but also different analytical choices in the processing of the fMRI data. While variability across participants or across acquisition instruments have been extensively studied in the neuroimaging literature, the root causes of analytical variability remain an open question. Here, we share the \textit{HCP Multi-Pipeline dataset} which provides the resulting statistic maps for 24 typical fMRI pipelines on 1,080 participants of the HCP Young Adult dataset. We share both individual and group results - for 1,000 groups of 50 participants - over 5 motor contrasts. We hope that this large dataset covering  a wide range of analysis conditions will provide new opportunities to study analytical variability in fMRI. 
\end{abstract}
\begin{document}

\flushbottom
\maketitle

\thispagestyle{empty}

\section*{Background \& Summary}
% A compléter 
Neuroimaging data, such as functional Magnetic Resonance Imaging (fMRI), can be used for a wide range of applications, including disease diagnosis~\cite{yin_deep_2022} or brain decoding (\textit{i.e.} identifying stimuli and cognitive states from brain activities)~\cite{firat_deep_2014}. But the workflows used to analyze these data are highly complex and flexible. Different tools and algorithms were developed over the years, leaving researchers with many possible choices at each step of an analysis~\cite{carp_plurality_2012}. A given series of operations performed on raw fMRI data is referred to as a `pipeline'. Task-fMRI analysis pipelines are composed of three high-level stages (preprocessing, subject-level analysis and group-level analysis). Each stage follows a base architecture with multiple processing steps: this sequence can be customized by the addition or removal of specific steps or modified by using different algorithms or set of parameters. Several software packages are also available to run task-fMRI analysis pipelines, for instance SPM~\cite{penny_statistical_2011}, FSL~\cite{jenkinson_fsl_2012} and AFNI~\cite{cox_afni_1996}, which are the most commonly used. 

In the Neuroimaging Analysis Replication and Prediction Study (NARPS)~\cite{botvinik2020variability}, 70 research teams were asked to analyze the same task-fMRI dataset with their favorite pipeline to answer the same 9 binary research questions investigating the activation of a particular brain region during a specific cognitive task. Each team used a different pipeline, illustrating perfectly how researchers can have different practices to analyze task-fMRI data. In the end, results varied in terms of final activation maps and conclusion to hypotheses. This phenomenon calls for a better understanding of the pipeline-space to try to identify the cause of the observed differences amongst the final results, and their impact on research findings. For instance, in a recent paper~\cite{demidenko_impact_2024}, authors explored whether some specific choices in the pipeline (in particular for the first-level statistical modelling) could improve the reliability of fMRI results, showing that specific choices for contrasts and model parameterization meaningfully improved test–retest reliability, both at the individual and group-level. 

The pipeline-space is especially large~\cite{carp_secret_2012} and challenging to explore due to its interaction with other properties of a dataset: for instance, with sample size and sampling uncertainty~\cite{klau2023comparing} or with the research question~\cite{botvinik2020variability}.  However, due to the high computational cost of storing and analyzing task-fMRI data, recent studies investigating analytical variability in neuroimaging focused on a restricted number of participants (N=108, N=30, N=15, and N=10 respectively for~\cite{botvinik2020variability, li2021moving, carp_plurality_2012,xu_guide_2022}) and cognitive tasks (one paradigm for~\cite{botvinik2020variability,carp_plurality_2012} with respectively k=9 and k=1 contrasts and use of resting-state fMRI for~\cite{li2021moving,xu_guide_2022}).

Multiple efforts for collecting datasets with larger number of participants and tasks have arisen in the field of neuroimaging in the past 10 years with for instance the Human Connectome Project (HCP)~\cite{hcp_overview} or the UK Biobank~\cite{sudlow_uk_2015,miller_multimodal_2016}. In particular, the HCP Young Adult most recent releases provide task-fMRI data for more than 1,000 participants and for different tasks and cognitive processes. These data are also available as minimally processed versions, \textit{i.e.} preprocessed using a common pipeline chosen by the HCP collaborators\cite{GLASSER2013105}. In brief, this pipeline consists in the following steps: removal of spatial distortions, volumes realignment to correct for subject motion, registration of the functional volumes to the structural one, bias field reduction, normalization to a global mean and masking using a structural brain mask computed in parallel. 

A set of group-level statistic maps of the HCP Young Adult have also been made publicly available (see NeuroVault Collection 457~\cite{neurovault_collection_457} and  corresponding publication~\cite{VANESSEN201362}). These were obtained using data from a subset of the participants (68 subjects scanned during the first quarter (Q1) of Phase II data collection. Z-scored statistic maps are available for all base contrasts (23 different contrasts) and were computed using a single analysis pipeline. This is beneficial for studying individual differences and contrasts, but it does not allow for analytical variability studies, for which multiple pipelines are needed, or to perform other analyses such as group-level analyses that could be used to explore the effect of sampling uncertainty or sample size. 

Statistic maps published during the NARPS study~\cite{botvinik2020variability} are also publicly available on NeuroVault, with one collection per team. For each of the 70 teams, 9 group-level statistic maps are shared (one per research hypothesis) based on two groups of N=54 participants. Additionally, for a limited number of teams (K=4), subject-level contrast maps are also available. The pipeline space studied in this dataset is unconstrained since teams were instructed to use their usual pipelines to analyze the data. 

Here, we share the \textit{HCP Multi-Pipeline dataset}, composed of a large number of subject and group-level statistic maps and representing a non-exhaustive but controlled part of the pipeline space. Contrast and statistic maps are made available for the 5 contrasts of the motor task of the Human Connectome Project (HCP Young Adult) for the 1,080 participants of the S1200 release, obtained with 24 analysis pipelines that differ on a predefined set of parameters as typically used in the literature. We also provide group-level contrast and statistic maps for 1,000 randomly sampled groups of 50 participants for each pipeline and contrast. 

While solutions have been proposed to standardized fMRI preprocessing (\textit{e.g.} fMRIprep~\cite{esteban_fmriprep_2019}), practitioners still face multiple choices regarding first-level statistical analyses. Here, we focus on a set of parameters that often vary across pipelines and this even when standardized preprocessing are used: smoothing kernels, HRF modelling and the inclusion/exclusion of motion regressors as nuisance covariates. Group-level statistical analyses were performed uniformly for all pipelines. 

\section*{Methods}

\subsection*{Raw Data: the Human Connectome Project}
This work was performed using data from the Human Connectome Project Young Adult~\cite{hcp_overview}. Written informed consent was obtained from participants and the original study was approved by the Washington University Institutional Review Board. We agreed to the Open Access Data Use Terms~\cite{hcp_dua}.

The HCP Young Adult aimed to study and share data from young adults (ages 22-35) from families with twins and non-twin siblings, using a protocol that included structural and functional magnetic resonance imaging (MRI, fMRI), diffusion tensor imaging (dMRI) at 3 Tesla (3T) and behavioral and genetic testing. The S1200 release includes behavioral and 3T MR imaging data from 1206 healthy young adult participants (1113 with structural MR scans) collected between 2012 and 2015. 

Unprocessed anatomical T1-weighted (T1w) and task-fMRI data~\cite{hcp_fmri_1,hcp_fmri_2,hcp_fmri_3,hcp_fmri_4} were used in this work. The task-fMRI data includes seven tasks, each performed in two separate runs. Among these tasks, we selected data from the motor task in which participants were presented with visual cues asking them to tap their fingers (left or right), squeeze their toes (left or right) or move their tongue. This task is the simplest one of the tasks performed in the study, and the protocol associated with this task is very standard and robust. We used unprocessed data for the $N=1080$ participants who completed this task. 

\subsection*{Analyses pipelines}\label{sec:analysispipeline}
Multiple preprocessing and first-level analyses were performed on the task-fMRI data, giving rise to 24 different analysis pipelines. These pipelines differed in 4 parameters: 
\begin{itemize}
    \item Software package: SPM (Statistical Parametric Mapping, RRID: SCR\_007037)~\cite{penny_statistical_2011} or FSL (FMRIB Software Library, RRID: SCR\_002823)~\cite{jenkinson_fsl_2012}.
    \item Smoothing kernel: Full-Width at Half-Maximum (FWHM) was equal to either 5mm or 8mm. 
    \item Number of motion regressors included in the General Linear Model (GLM) for the first-level analysis: 0, 6 (3 rotations, 3 translations) or 24 (the 6 previous regressors + 6 derivatives and the 12 corresponding squares of regressors).
    \item Presence (1) or absence (0) of the derivatives of the Hemodynamic Response Function (HRF) in the GLM for the first-level analysis. Only the temporal derivatives were added in FSL pipelines and both the temporal and dispersion derivatives were for SPM pipelines.
\end{itemize}

In the following, we will denote the pipelines by ``software-FWHM-number of motion regressors-presence of hrf derivatives''. For instance, pipeline with software FSL, smoothing with a kernel FWHM of 8mm, no motion regressors and no hrf derivatives will be denoted by ``fsl-8-0-0''. 

All pipelines were implemented using Nipype version 1.6.0 (RRID: SCR\_002502)\cite{gorgolewski_nipype_2017}, a Python project that provides a uniform interface to existing neuroimaging software packages and facilitates interaction between these packages within a single workflow. All pipelines scripts are available online at \url{https://gitlab.inria.fr/egermani/hcp_pipelines} and archived on Software Heritage: \href{https://archive.softwareheritage.org/swh:1:snp:17870c3d782aa25a7ffdd6165fe27ce6eac6c90b;origin=https://gitlab.inria.fr/egermani/hcp_pipelines}{swh:1:snp:17870c3d782aa25a7ffdd6165fe27ce6eac6c90b}

\subsubsection*{Computing environment}
To limit the variability induced by different computer environments and versions of the software packages, we used NeuroDocker (RRID: SCR\_017426)\cite{neurodocker} to generate a custom Docker image. To build this image, we chose NeuroDebian\cite{halchenko_open_2012} and installed the following software packages: FSL version 6.0.3 and SPM12 release r7771. 
To install Python and Nipype, we created a Miniconda3 environment with Python version 3.8 and multiple packages, such as Nilearn~\cite{abraham_machine_2014} (RRID: SCR\_001362), Nipype and NiBabel (RRID: SCR\_002498)~\cite{brett_nipynibabel_2020}. This docker image is available at \url{https://hub.docker.com/repository/docker/elodiegermani/open_pipeline/general} and the command to generate the DockerFile can be found in the README included in the codebase (see links above). 

\subsubsection*{Preprocessing}
Preprocessing consisted of the following steps for all pipelines: spatial realignment of the functional data to correct for motion, coregistration of realigned data towards the structural data, segmentation of the structural data, non-linear registration of the structural and functional data towards a common space and smoothing of the functional data. Depending on the software package used, these steps were performed in a different order, following the default behavior of the software package. 

In SPM, for each participant, functional data were first spatially realigned to the mean volume using the ``Realign: Estimate and Reslice'' function with default parameters (quality of 0.9, sampling distance of 4 and a smoothing kernel, 2nd degree B-spline interpolation and no wrapping). Realigned functional data were then coregistered, with the ``Coregister: Estimate'' function, to the anatomical T1w volume acquired for the participant using Normalized Mutual Information. In parallel, we segmented the different tissue classes of the same anatomical T1w volume using the ``Segment'' function. The forward deformation field provided by the segmentation step was used to normalize the functional data to a standard space (MNI)(``Normalize: Write'' function) with a voxel size of 2mm and a 4th degree B-spline interpolation. Normalized functional data were then smoothed with a Gaussian kernel with different FWHM values depending on the pipeline (5 or 8mm).

In FSL, we reproduced the preprocessing steps used in FEAT~\cite{feat_1}. Functional data were realigned to the middle functional volume using MCFLIRT. Brain extraction was applied with BET and we masked the functional data using the extracted mask. We smoothed each run using SUSAN with the brightness threshold set to 75\% of the median value for each run and a mask constituting the mean functional. Different values were used for the FWHM of the smoothing kernel depending on the pipeline. We also performed temporal highpass filtering on the functional data with a thresholding value of 100 seconds. In parallel, we computed the transformation matrix to register functional data to anatomical and standard space (MNI) using linear (FLIRT function) and non-linear registration (FNIRT function). Contrary to SPM, the first-level statistical analysis is performed on the smoothed data in subject-space. Only the transformation matrix was computed at this stage, using boundary-based registration and applied on the contrast maps output after the statistical analysis. 

\subsubsection*{First level statistical analyses}

To obtain the contrast maps of the different participants and contrasts, we modeled the data using a GLM. Each event was modelled using the onsets and durations provided in the event files of the HCP Young Adult dataset. Six regressors were modeled: cue (which represent the visual cues), right hand, right foot, left hand, left foot and tongue. Each condition was convolved with the canonical HRF. For both SPM and FSL pipelines, we used the Double Gamma HRF (default in SPM).  

Different numbers of motion regressors (0, 6 or 24) were included in the design matrix to regress out motion-related fluctuations in the BOLD signal. The modelling of the HRF also varied: Double Gamma HRF with or without derivatives (time+dispersion for SPM and time for FSL). 

In SPM, temporal autocorrelations in the BOLD signal timeseries were accounted for by highpass filtering with a 128 seconds filter cutoff and modelling of serial correlation using an autoregressive model of the first order (AR(1)). In FSL, highpass filtering was already performed during preprocessing with a 100 seconds filter cutoff, modelling of serial correlation was also performed using an AR(1) model. Model parameters were estimated using a Restricted Maximum Likelihood approach for both SPM and FSL software packages. Subject-level contrast maps were computed and saved for 5 contrasts (right hand, right foot, left hand, left foot and tongue) for each participant. In the end, for each of the 24 pipelines, we had 5,400 contrast and statistic maps (5 contrasts for each of the 1,080 participants). These maps constituted the subject-level dataset. Figure \ref{fig:fig1}(A) presents the statistic maps for the contrast \textit{right-hand} obtained with the different pipelines for a representative subject.

\subsubsection*{Second level statistical analyses}
\begin{figure}[p]
    \centering
    {\vspace{-1cm}
    \includegraphics[width=18cm]{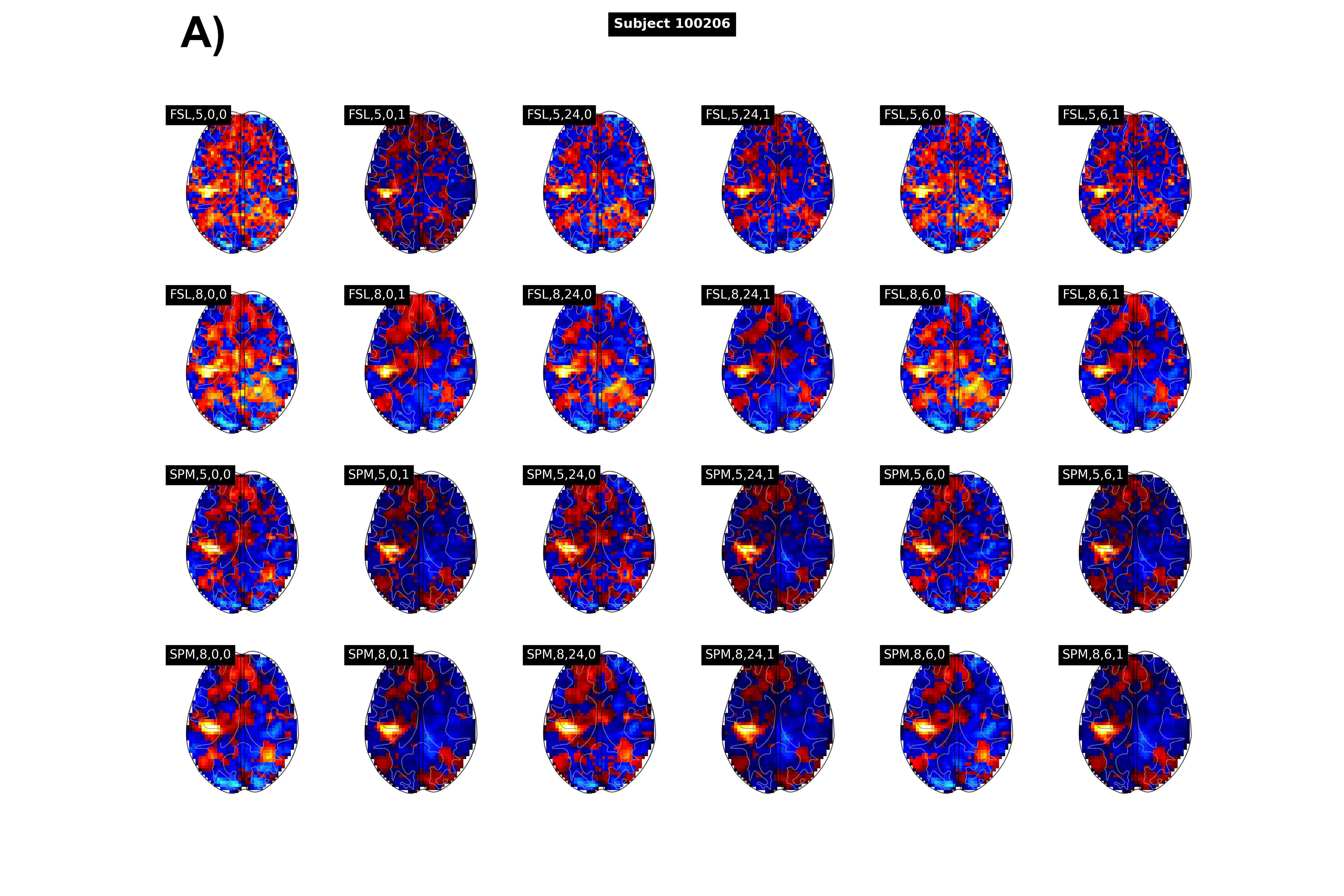}
    \vspace{-1cm}
    \includegraphics[width=18cm]{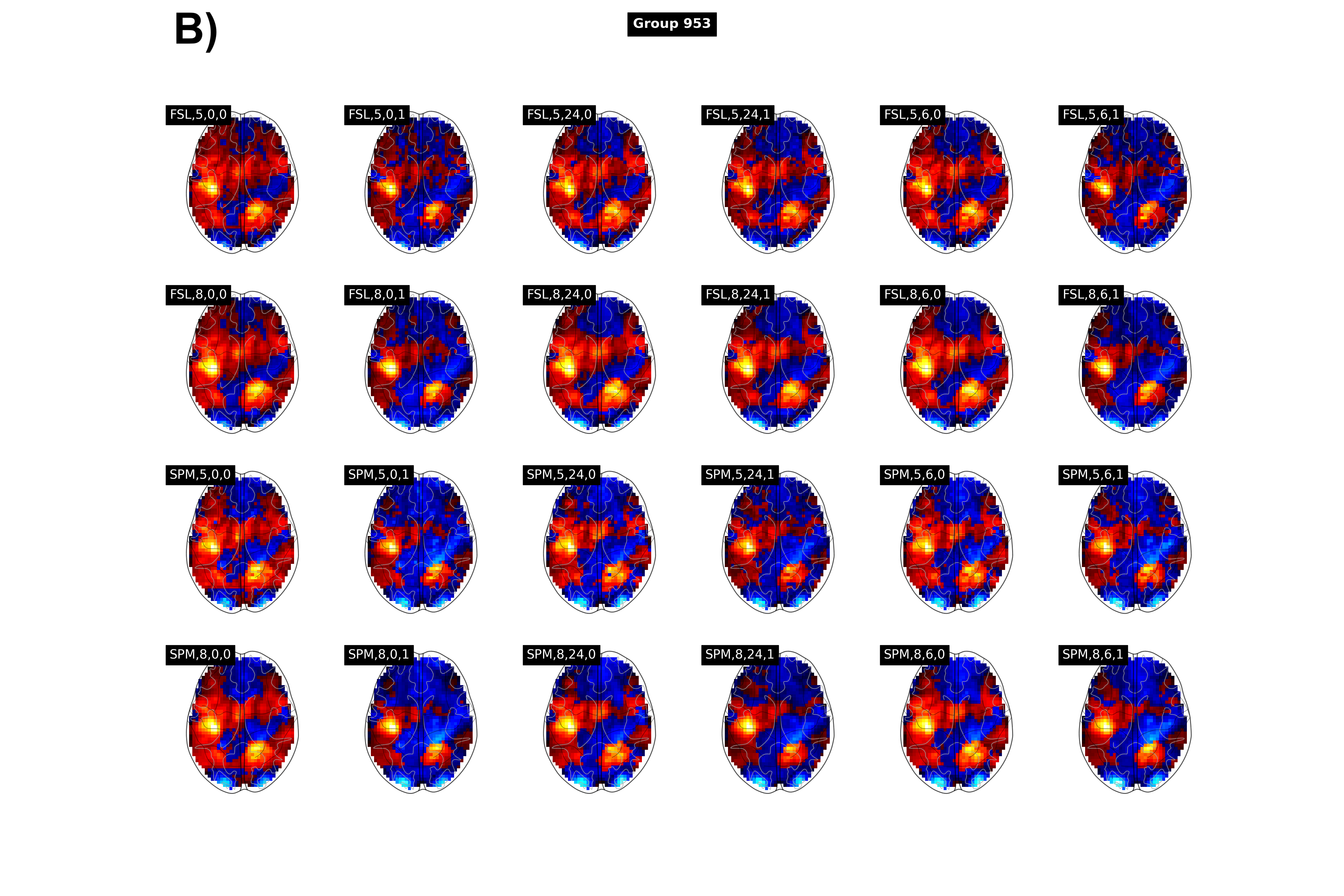}}
    \caption{Example of subject (A) and group-level (B) statistic maps obtained for subject 100206 and group 953 for each pipeline for the contrast \textit{right-hand}. Pipelines are denoted by \textit{<software>,<FWHM>,<motion regressors>,<hrf derivatives>}.}
    \label{fig:fig1}
\end{figure}

Group-level statistical analyses were performed using the contrast maps obtained with the different first-level analyses pipelines. 1,000 groups of 50 participants were randomly sampled among the 1,080 participants. For each analysis pipeline, we performed one sample t-tests for each group and each contrast in SPM. We purposely used the same second-level analysis method and software for all pipelines in order to focus on first-level analysis differences. 
For each of the 24 pipelines, the group-level dataset was thus composed of 5,000 contrast maps and statistic maps (5 contrasts for each on the 1,000 groups). Figure \ref{fig:fig1}(B) presents the statistic maps obtained with the different pipelines for one group for the contrast \textit{right-hand}.

\section*{Data Records}
The data are accessible on Public nEUro~\cite{publicneuro} at \url{https://publicneuro-catalogue.netlify.app/dataset/PN000003:%20HCP%20multipipelines/V1}. The dataset is organized according to the BIDS specification~\cite{gorgolewski_brain_2016}, but to date, there is no established specification for organizing datasets with statistic maps for different pipelines and different processing levels (here, subject and group-levels derivatives). We used a modlisation that is currently under development in the BIDS Extension Proposal: BEP041 - Statistical Model Derivatives. 

Briefly, the dataset contains one folder per pipeline (\textit{i.e.} 24 folders in total). Each pipeline folder is composed of two subfolders respectively named ``node-L1'' and ``node-L2'' for subject-level and group-level derivatives. The ``node-L1'' folder includes subject-level folders, named after the BIDS specifications, and the ``node-L2'' folder is composed of group-level folders, named `group-X' depending on the group number. Group and subject-level folders contain the contrast and statistic maps with the following specification for contrast maps: `sub/group-<ID>\_task-<task>\_space-<space>\_contrast-<contrast>\_stat-effect\_statmap.nii.gz', and T-statistic maps: `sub/group-<ID>\_task-<task>\_space-<space>\_contrast-<contrast>\_stat-t\_statmap.nii.gz'. 
 
\section*{Technical Validation}
To assess the quality of the statistic maps, we checked that all contrasts led to an activation of the primary motor area (Primary Motor Cortex M1). 

\begin{figure}[htb!]
    \centering
    \includegraphics[width=18cm]{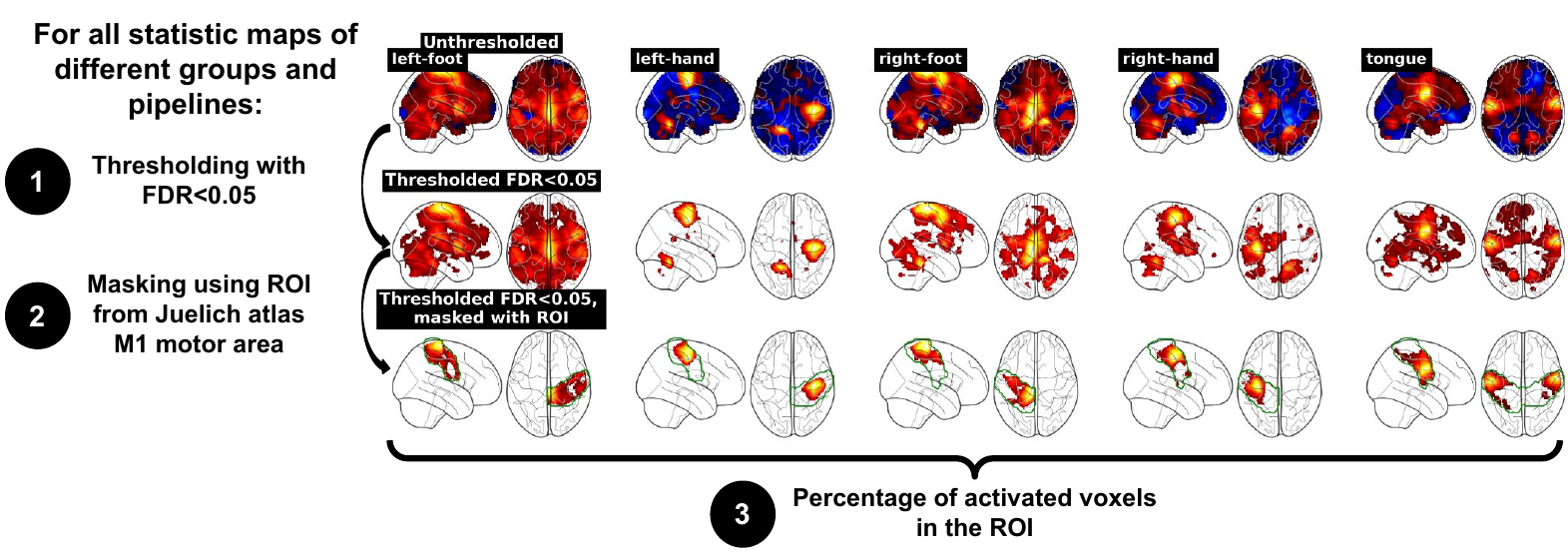}
    \caption{Workflow of technical validation of statistic maps. We thresholded the statistic maps of each group, each pipeline and each contrast using a FDR-corrected voxelwise $p<0.05$ and masked the thresholded maps using the ROI of the Primary Motor Cortex from Juelich atlas. Then, we computed the percentage of activated voxels inside this ROI.}
    \label{fig:fig2}
\end{figure}

As described in Fig.~\ref{fig:fig2}, we looked at the significant activations inside the Primary Motor Cortex (M1) of the brain for the statistic maps of each group, each contrast and each pipeline. Our group-level statistic maps were thresholded using a False Discovery Rate (FDR) corrected voxelwise p-value of \(p<0.05\) and masked using the probabilistic Juelich Atlas~\cite{amunts_julich-brain_2020} available in Nilearn. We selected the region of interest (ROI) corresponding to the Primary Motor Cortex M1, Brodmann Area 4. Depending on the contrast, both left and right hemisfer's ROI (``tongue''), only the left hemisfer (``right hand'' or ``right foot'') or only the right hemisfer (``left hand'' or ``left foot'') ROI were selected, to focus on controlateral activations in the motor cortex. 

For each map, we computed the percentage of activation inside the Primary Motor Cortex, which is the percentage of voxels of the ROI that are activated, \textit{i.e.}:
\begin{equation}\label{eq1}
    Percentage \ of \ Activation = \frac{N_{activated\ voxels}}{N_{total\ voxels}} \times 100
\end{equation} 
where $N_{activated\ voxels}$ is the number of activated voxels in the ROI and $N_{total\ voxels}$ is the total number of voxels in the ROI. \\

\begin{figure}[htb!]
    \centering
    \includegraphics[width=15cm]{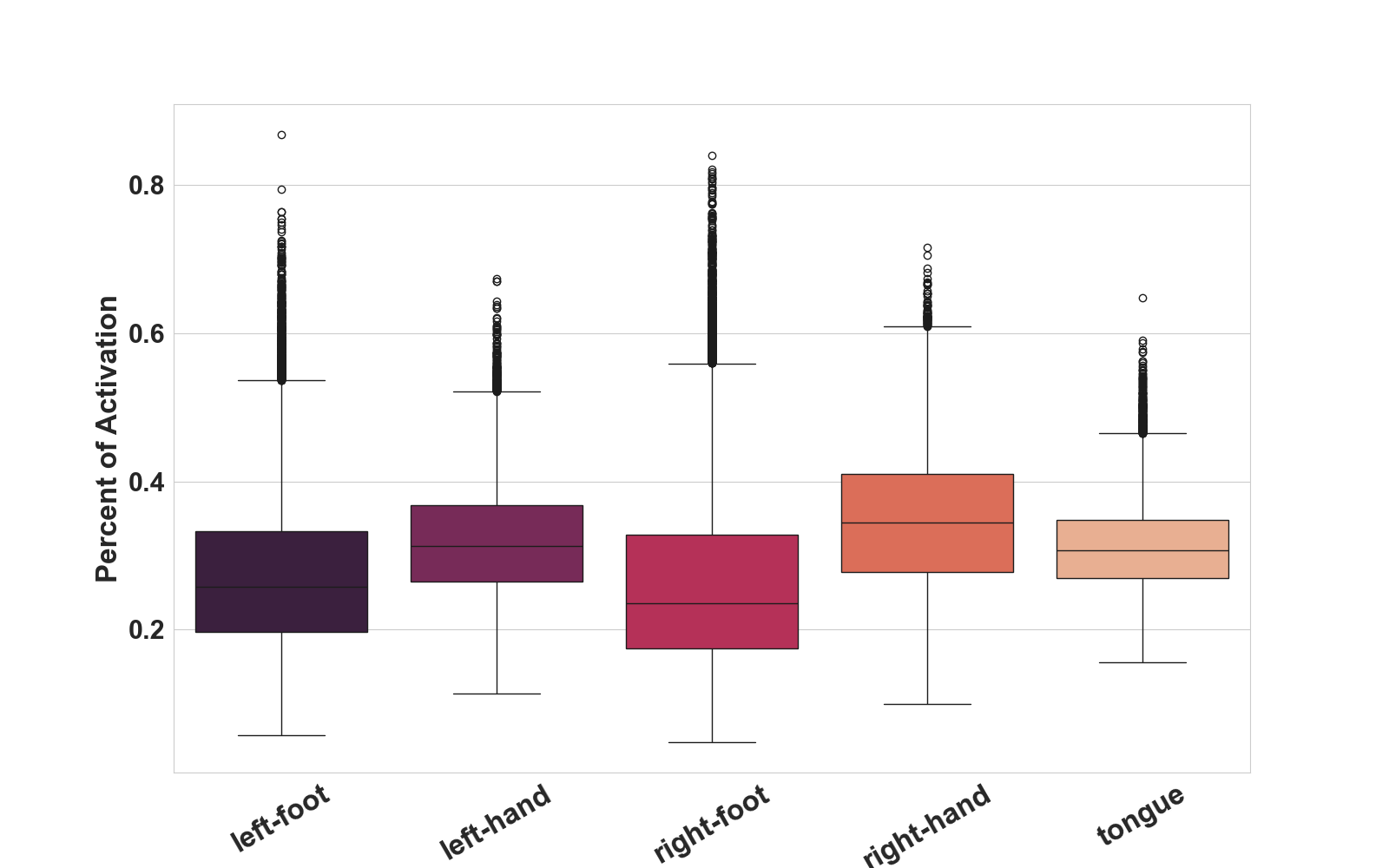}
    \caption{Distribution of mean Percentage of Activation inside the Primary Motor Cortex across all groups and pipelines for the five contrasts under study.}
    \label{fig:fig3}
\end{figure}

\begin{figure}[htb!]
    \centering
    \includegraphics[width=12cm]{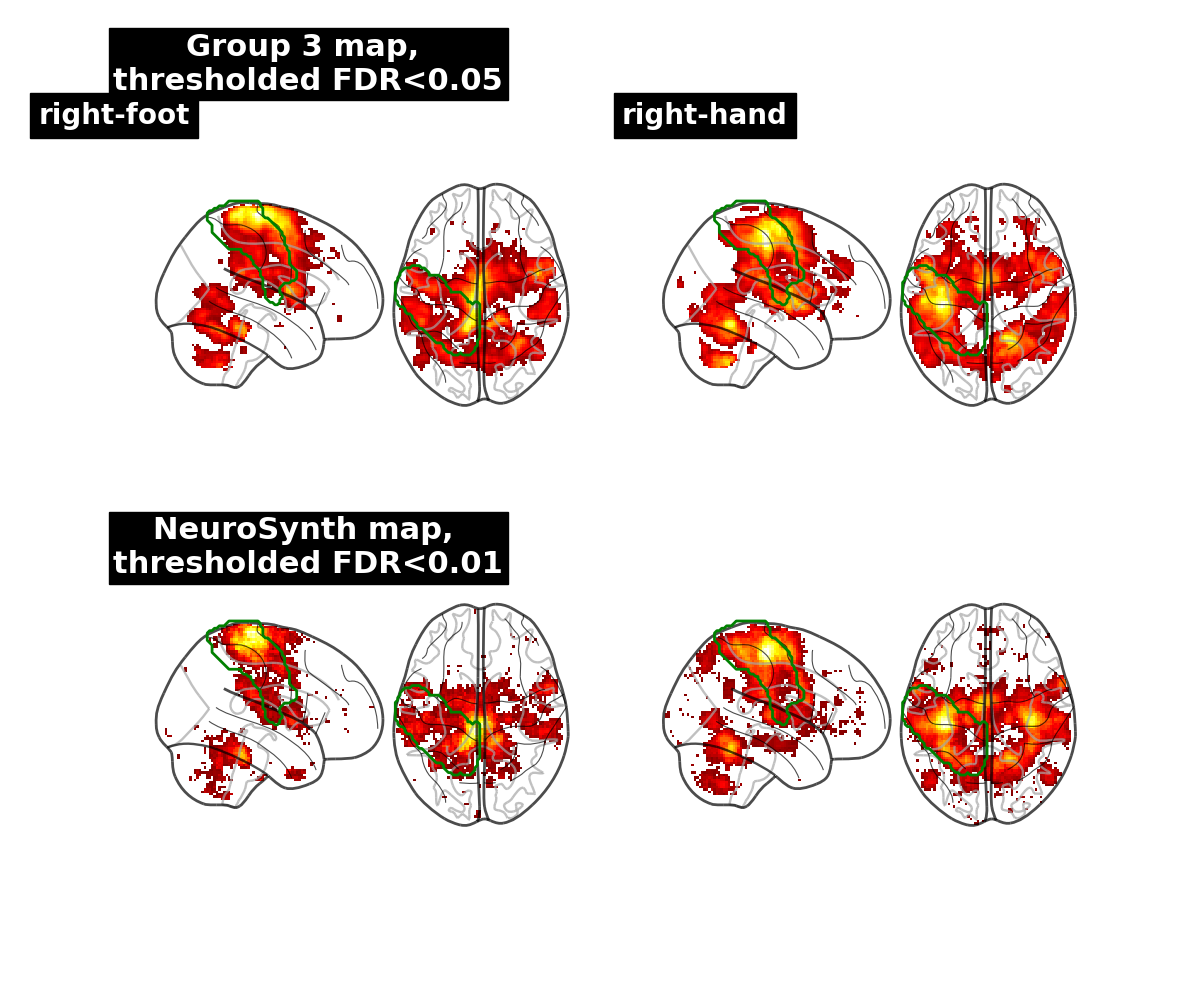}
    \caption{Thresholded statistic maps for contrasts \textit{right foot} (right) and \textit{right hand} (left) for group-level analysis of group~3 with pipeline spm-5-0-0 (upper). Percentage of Activation inside the Primary Motor Cortex were respectively 0.34 and 0.41 for the contrasts \textit{right foot} and \textit{right hand}. NeuroSynth activation maps corresponding to the forward inference of the ``hand'' and ``foot'' paradigms (lower). Green borders correspond to the motor area ROI.} 
    \label{fig:fig4}
\end{figure}

Fig.~\ref{fig:fig3} represents the distribution of percentages of activation inside the Primary Motor Cortex per contrast for all groups and all pipelines. Results were slightly different across contrasts: mean percentages of activation were all between 20 and 40\% but those of contrasts \textit{left-foot} and \textit{right-foot} were below those of \textit{right-hand}, \textit{left-hand} and \textit{tongue}. When looking at the activations of different contrasts in the ROI for one of our group-level statistic maps (see Fig.~\ref{fig:fig4}), we could see that the activations detected for the \textit{right-foot} contrast seemed widespread with a small area of high activation. For the \textit{right-hand} contrast, the high activation area was larger and covered nearly the entire ROI. This observation was consistent with the literature~\cite{homonculus,yarkoni_large-scale_2011} and with statistic maps obtained from NeuroSynth~\cite{yarkoni_large-scale_2011} (RRID:SCR\_006798) in which the identified area of activation of the foot was smaller than the hand one (see Fig.~\ref{fig:fig4}). In the Primary Motor Area, the statistic maps of the foot contrasts thus have less activated voxels.  Overall, the technical validation was successful. The goal of this quality check was to have a quick estimation of the accuracy of the statistic maps to represent the task performed, thus we chose to define a single ROI covering the entire motor area. 

\begin{figure}[htb!]
    \centering
    \includegraphics[width=15cm]{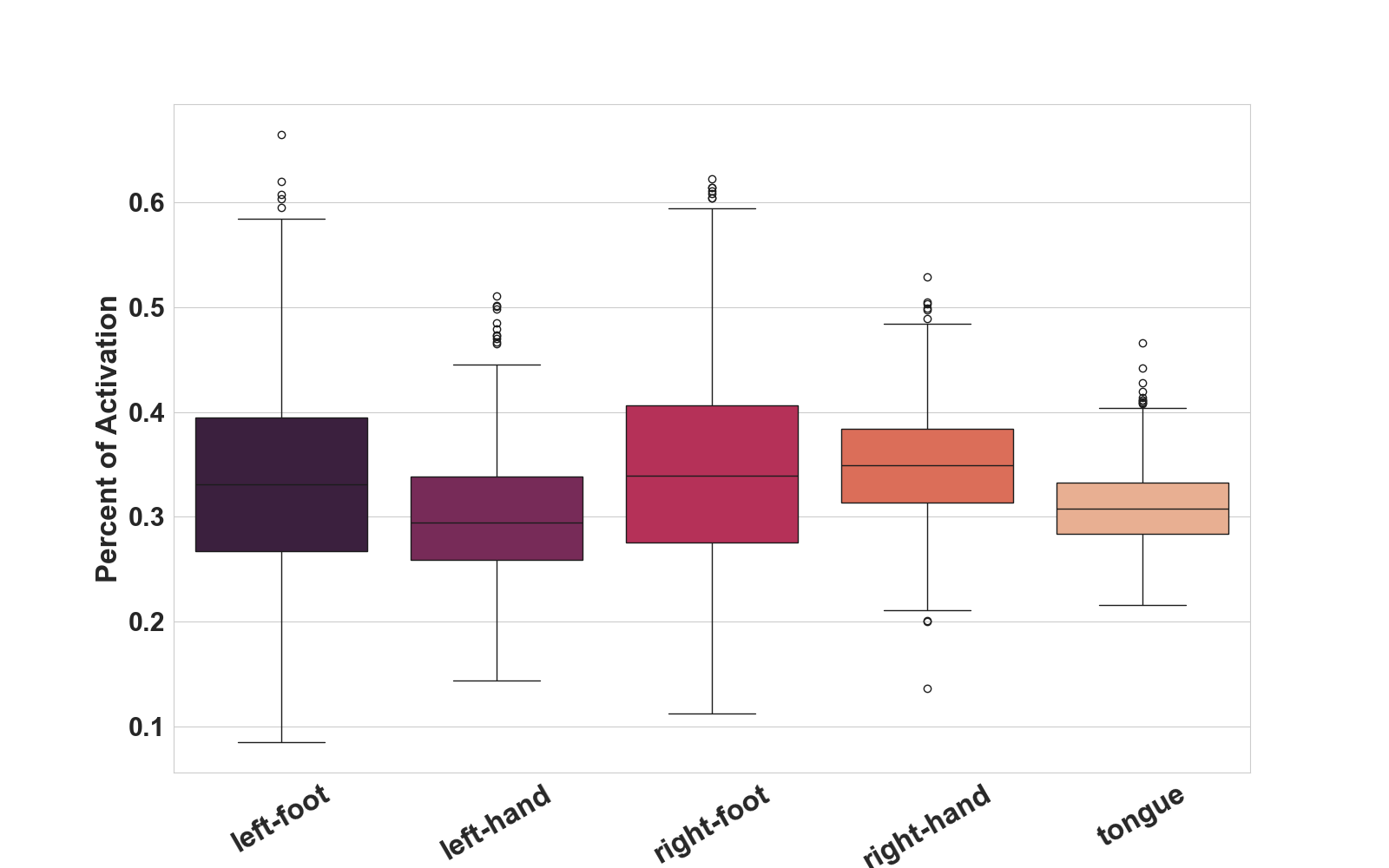}
    \caption{Distribution of Percentage of Activation inside the Primary Motor Cortex for all group-level statistic maps for pipeline ``spm-5-0-0'' in the different contrast maps.}
    \label{fig:fig5}
\end{figure}

We observed consistent metrics across pipelines, with high percentages of activation for hand and tongue contrasts and lower ones for foot contrasts. An example of the distribution of percentage of activations for all group maps of each contrast is shown in Fig.~\ref{fig:fig5} for the pipeline ``spm-5-0-0''. The percentage of activation computed for all groups, pipelines and contrasts are available in Supplementary Files. 

\section*{Usage Notes}
The \textit{HCP Multi-Pipeline dataset} provides researchers with a re-usable dataset of fMRI contrast and statistic maps. The data is accessible on Public nEUro~\cite{publicneuro} at \url{https://publicneuro-catalogue.netlify.app/dataset/PN000003:%20HCP%20multipipelines/V1}.. 

This dataset brings together a wide range of analysis conditions, covering many aspects of inter-subject, inter-groups, inter-contrasts and inter-pipelines variability. Data from 1,080 participants were used to form 1,000 different groups of 50 participants, and 5 contrasts were analyzed with 24 different pipelines, which offers an unprecedented opportunity to explore analytical variability, understand its causes and how it may interact with other sources of variability. 

Analytical variability is not limited to neuroimaging and has been studied in many other disciplines~\cite{hoffmann2021multiplicity}, such as psychology~\cite{simmons_false-positive_2011} or software engineering~\cite{alferez_modeling_2019}.  These different fields have brought solutions to explore and handle analytical variability. These techniques have begun to be used in neuroimaging, with, for instance, the implementation of continuous integration, a software engineering technique, to facilitate the reproducibility of neuroimaging computational experiments~\cite{sanz-robinson_neuroci_2022} or multiverse analyses that help to find the most efficient pipelines depending on the data and the goal of the study~\cite{dafflon2022guided}.

By sharing directly the results obtained from different analysis strategies, we hope to facilitate the use of these data by researchers from other fields, that could apply their own methods to help explore the neuroimaging analytical space. For instance, this dataset could be use to extend the study from~\cite{demidenko_impact_2024} to other tasks. Using the code provided to create the pipelines, other researchers could be able to enhance this dataset with other combinations of parameters, giving rise to other pipelines, or could apply the pipelines to other participants, groups or contrasts.

\section*{Code availability}
All codes (analysis pipelines and technical validation) were executed in Python v3.8. The executions require the installation of SPM and FSL software packages. To facilitate reproducibility, we provide a NeuroDocker image that can be pulled from Dockerhub and that contains all necessary software packages. The Docker image is available at: \url{https://hub.docker.com/r/elodiegermani/open_pipeline}. 
Python scripts to create and run the pipelines as well as to perform technical validation are available publicly at \url{https://gitlab.inria.fr/egermani/hcp_pipelines} and archived on Software Heritage at \href{https://archive.softwareheritage.org/swh:1:snp:17870c3d782aa25a7ffdd6165fe27ce6eac6c90b;origin=https://gitlab.inria.fr/egermani/hcp_pipelines}{swh:1:snp:17870c3d782aa25a7ffdd6165fe27ce6eac6c90b}.

\bibliography{sample}

\section*{Acknowledgements}

This work was partially funded by Region Bretagne (ARED MAPIS) and Agence Nationale pour la Recherche for the programm of doctoral contracts in artificial intelligence (project ANR-20-THIA-0018).
Data used in the HCP dataset were provided by the Human Connectome Project, WU-Minn Consortium (Principal Investigators: David Van Essen and Kamil Ugurbil; 1U54MH091657) funded by the 16 NIH Institutes and Centers that support the NIH Blueprint for Neuroscience Research; and by the McDonnell Center for Systems Neuroscience at Washington University.

We thank the Inria open data team and in particular Jozefina Sadowska and Maud Medves who made it possible to share this dataset. We also thank our data protection officers, security, finance and legal teams and in particular Caroline Surquain for her continued support throughout the process. Finally we would like to thank Remi Gau and Chris Markiewicz for their guidance to format the dataset in compliance with BIDS.

\section*{Author contributions}
\textbf{Elodie Germani}: Conceptualization, Methodology, Software, Formal analysis, Investigation, Data Curation, Writing - Original Draft.
\textbf{Elisa Fromont}: Supervision, Writing - Review \& Editing. 
\textbf{Pierre Maurel}: Supervision, Writing - Review \& Editing. 
\textbf{Camille Maumet}: Supervision, Writing - Review \& Editing. 

\section*{Competing interests} 
Camille Maumet is an Editorial Board Member for Scientific Data, handling manuscripts for the journal and advising on its policy and operations.
The other authors declare no competing interests.

\end{document}